Efficient light coupling from integrated single-mode waveguides to supercollimating photonic crystals on silicon-on-insulator platforms

K. Vynck[a)], E. Centeno, M. Le Vassor d'Yerville, and D. Cassagne

Groupe d'Etude des Semiconducteurs, UMR 5650 CNRS-Université Montpellier II, CC074, Place E. Bataillon, 34095 Montpellier Cedex 05, France

We propose a practical and efficient solution for the coupling of light from integrated single-mode waveguides to supercollimating planar photonic crystals on conventional silicon-on-insulator platforms. The device consists of a rib waveguide, designed to sustain spatially extended single-modes and matched to a supercollimating photonic crystal, which has been truncated at its boundary to improve impedance matching between the two photonic components. Three-dimensional simulations show transmission efficiencies up to 96 % and reflections below 0.2 % at wavelengths close to 1.55 μm. This approach constitutes a significant step toward the integration of supercollimating structures on photonic chips.

In the late 1980's, a new branch of electromagnetism emerged with the observation that periodic arrangements of dielectric media, or Photonic Crystals (PhCs)[1], could be used to control the emission and propagation of light in three-dimensional (3D) space. For many years, researchers and engineers made use of the photonic band gap exhibited by certain PhCs to confine the light to structural defects[2]. More recently, it appeared that the richness of the PhCs dispersive properties could provide a multitude of novel optical effects such as ultrarefraction[3], negative refraction[4] or supercollimation[5], and therefore bring up new functionalities to PhCs. Supercollimation in particular gives the possibility to propagate light over the centimetre scale without the use of structural waveguides. Recent studies report the possibility to realize devices such as optical routers, multiplexers or polarization splitters, by incorporating linear defects in periodic structures[6,7] or alternatively by using graded PhCs[8]. In fact, supercollimating PhCs are the building block of many promising applications and it seems very likely that they will play a key role in future telecommunications technologies.

The proper integration of supercollimating PhCs on a photonic chip implies that light has to be efficiently transmitted throughout the whole device using integrated and single-mode waveguides. The coupling of light to supercollimating PhCs however remains a technological challenge. First, the incident beam has to excite very specific Bloch modes in the PhC[9], imposing some requirements on the design of the excitation waveguide, in addition to the single-mode condition necessary to insure proper light signal transmission between photonic components. To our knowledge, the coupling techniques[10-14] that have been proposed or exploited up to now do not use integrated single-mode waveguides, in spite of their necessity, simply because no practical and efficient solution has been found. Second, impedance mismatch at the waveguide-PhC interface causes reflections and thus Fabry-Pérot oscillations in finite-size structures that may be inconvenient in practical applications. Although it has been shown that the PhC boundary is of crucial importance for improving the coupling of light to PhCs[15,16], such an approach has not been applied yet to supercollimating PhCs in particular. In this Letter, we report an efficient coupling between an integrated single-mode waveguide and a supercollimating PhC, both realizable on a single photonic platform and designed to operate in the telecommunications range of wavelengths. We design the excitation waveguide according to the supercollimating PhC beam propagation requirements and then optimize the PhC boundary to improve impedance matching at the waveguide-PhC interface.

Due to the wide use of silicon (Si) in present electronic devices, the pre-eminent solution for integrating photonics to future technologies is to use silicon-on-insulator (SOI) platforms. The high index contrast between Si and silica ($SiO_2$) makes it possible to design efficient planar PhCs and low-loss waveguides. Rib waveguides[17] in particular consist of a waveguide core surrounded by a partially etched Si layer, contrary to strip or ridge waveguides, where the surrounding Si layer if etched down to the $SiO_2$ layer. Rib waveguides have the capability of sustaining TE- and TM-like single-modes that exhibit a larger spatial extension than other SOI-based waveguides. As explained below, these spatially extended propagating modes are the key for a selective and efficient coupling of light to the PhC supercollimated modes. On an experimental point of view, both components (waveguide and PhC) can be fabricated on a single SOI wafer, which therefore makes the whole device easy to integrate on current photonic platforms. A 3D view of the device under study is given on Fig.1.

The PhC under consideration consists of a 2D square arrangement of air holes in a Si layer of refractive index 3.5 and thickness $t$=340 nm, deposited on a $SiO_2$ layer of refractive index 1.45 and thickness 700 nm that lies on a thick Si substrate. One should note that a thickness of 700 nm is sufficient for the $SiO_2$ layer below the guiding Si layer to prevent light from leaking down to the Si substrate[17]. The period $a$ of the lattice is 310 nm and the radius $r$ of the holes is 94 nm. These parameters were chosen to collimate light at wavelengths $\lambda$ close to 1.55 μm and insure its vertical confinement to the PhC slab by means of total internal reflection. The dispersive properties of PhCs can be obtained from the PhC Iso-Frequency Curves (IFCs) in reciprocal space, where the group velocity of a propagating mode is defined by $\mathbf{v}_g = \nabla_{\mathbf{k}} \omega(\mathbf{k})$ and its direction by the normal to the IFC. The PhC dispersion curves were computed with the Finite-Difference Time-Domain (FDTD) method in 3D using a freely available software package with subpixel smoothing for increased accuracy[18] and a filter diagonalization technique to extract the PhC resonant modes frequencies[19].

Figure 2 sketches the IFCs corresponding to the first TE-like mode of the PhC. Straight IFCs, which are composed by modes with nearly similar group velocity directions, appear along the ΓM direction for reduced frequencies ($a/\lambda$) between 0.19 and 0.21. The near-zero IFC curvature observed in this region lies on a reciprocal half-width $\Delta k$ of 0.05 ($2\pi/a$), corresponding to a minimal lateral half-width $\Delta x$ of $1.6a$=0.50 μm, according to the relation $\Delta x \cdot \Delta k \geq 1/2$. The lateral width of the incident beam should therefore be greater than or equal to $2\Delta x = 1.0$ μm to prevent some of the incident modes to lie outside the IFC straight area and thus to be dispersed in the PhC.

Single-modes with a lateral width greater than 1.0 μm can however not be sustained in conventional strip (or ridge) waveguides. Rib waveguides, on the contrary, have this capability, which makes them good candidates for an efficient coupling to supercollimating PhCs. As a proof of concept, we consider a rib waveguide with a total height $H$=340 nm (fixed by the planar PhC thickness $t$), an etch depth $d$=40 nm and a width $W$=700 nm. Figure 3 shows the corresponding TE-like single-mode propagating at a wavelength of 1.55 μm ($a/\lambda$=0.200 in the PhC) as calculated by a commercial finite-element software package[20]. The full-width at half-maximum of the mode at the depth where it spreads the most is about 1.2 μm, which is larger than the minimal lateral width of the supercollimated beam considered above (1.0 μm). This rib waveguide can therefore be used confidently as the excitation waveguide of our device.

In addition to this minimal lateral width condition, the light coupling efficiency also relies on the quality of the impedance matching at the coupling interface between the two photonic components. It is known that the PhC boundary can be optimized to improve impedance matching in a way to reduce undesirable reflections[15,16]. This optimization can be realized by truncating the PhC at different positions, as it has been done in negative refraction PhCs[21,22]. By a series of FDTD calculations, which for the sake of clarity have not been included in this Letter, we find that reflections are minimized when the supercollimating PhC is truncated in the middle of a column of holes along the ΓM direction of the PhC square lattice (see the inset of Fig. 4a).

The coupling efficiency at the waveguide-PhC interface is finally obtained by calculating the transmission between two rib waveguides (input and output) placed on each side of an 8 μm-long supercollimating PhC directed along the ΓM direction of the PhC square lattice. Simulations were performed in 3D with the FDTD method[18]. The spectra shown on Fig. 4a exhibit transmission efficiencies up to about 96 % and quasi-null reflections (below 0.2 %) at wavelengths close to 1.55 μm ($a/\lambda = 0.200$). The losses, which quantify the amount of light that is dispersed at the waveguide-PhC interfaces and within the PhC, go down to about 4 %. Such low losses, low reflections, and high transmission efficiencies therefore infer an excellent coupling efficiency at the waveguide-PhC interfaces, especially considering that the out-of-plane losses were also taken into account in the calculation. The Fabry-Pérot oscillations at wavelengths close to 1.55 μm are particularly weak, which demonstrates the excellent impedance matching obtained at the waveguide-PhC interfaces from the truncation of the PhC. The steady-state amplitude of the magnetic field at a wavelength of 1.55 μm given on Fig. 4b confirms the low reflections and dispersion experienced at the waveguide-PhC interfaces and within the PhC. One should finally note that at wavelengths where the IFC curvature is larger, light is more dispersed within the PhC, yielding an increase of the losses. Reflections however remain low, which shows that the coupling efficiency at the waveguide-PhC interface is excellent on the whole wavelength range of study.

To conclude, we showed that rib waveguides, due to their capability of sustaining spatially extended single-modes, are an efficient and easy-to-integrate way of coupling single-modes to supercollimating PhCs. As an example, we studied a particular photonic device with the FDTD and finite-element methods, both in 3D, and calculated transmission efficiencies as high as 96 %, and reflections below 0.2 % at wavelengths close to 1.55 μm. We improved impedance matching and thereby reduced undesirable reflections at the waveguide-PhC interface by truncating the PhC at its boundary. Current experimental techniques give the possibility to fabricate such structures on SOI platforms and we are confident that this approach will significantly improve the quality of future photonic devices based on supercollimation. In a more general view, it opens new perspectives for the coupling of light to extended planar PhCs and may also contribute to the integration of negative refraction, ultrarefraction or graded PhCs on photonic chips.

**Acknowledgements**

The authors acknowledge the CINES "Centre Informatique National de l'Enseignement Supérieur" for an allowance of computer time. This work is supported in part by the EU-NoE Project Nr 511616 PhOREMOST "NanoPhotonics to Realise Molecular Scale Technologies".

**Footnotes and references**

**Figures**

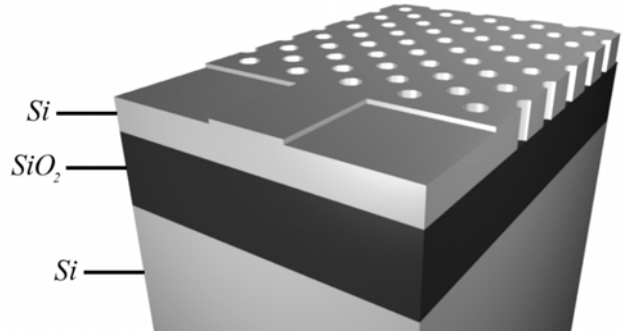

FIG. 1. Schematic of the waveguide-PhC interface on a SOI wafer, consisting of a patterned Si layer, deposited on a SiO$_2$ layer that lies on a thick Si substrate.

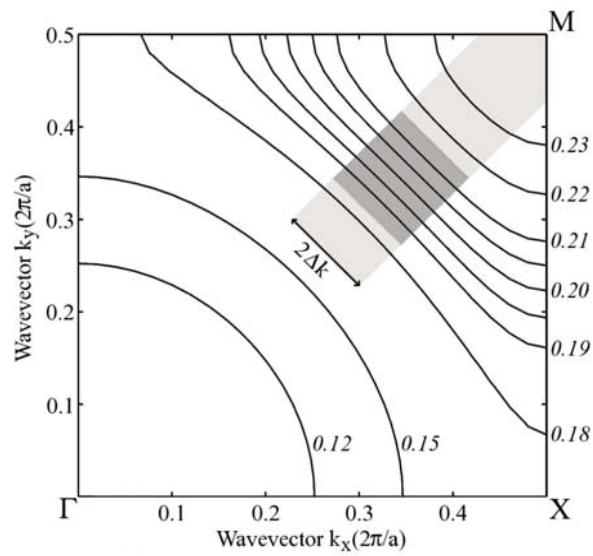

FIG. 2. IFCs of a PhC consisting of a square lattice of holes with radius $r$=94 nm and a lattice period $a$=310 nm, calculated with the FDTD method in 3D. The reduced frequencies are indexed on their respective IFC in units of $a/\lambda$. The dark gray-shaded area corresponds to the reciprocal full-width ($2\Delta k$) on which the IFCs exhibit a near-zero curvature.

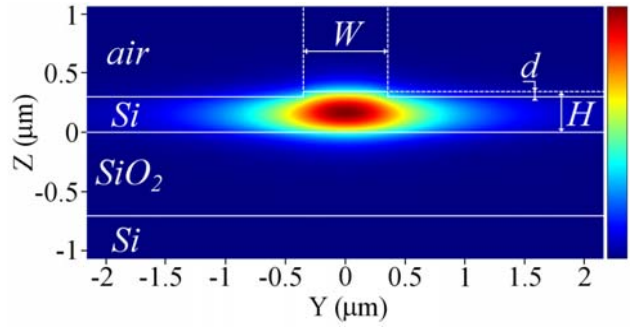

FIG. 3. (Color online) Amplitude of the magnetic field $H_z$ at a wavelength of 1.55 μm in a cross-sectional view of the rib waveguide, calculated with the finite-element method. The waveguide parameters are $H$=340 nm, $d$=40 nm and $W$=700 nm. The Si layer is deposited on a 700 nm-thick $SiO_2$ layer that lies on a thick Si substrate.

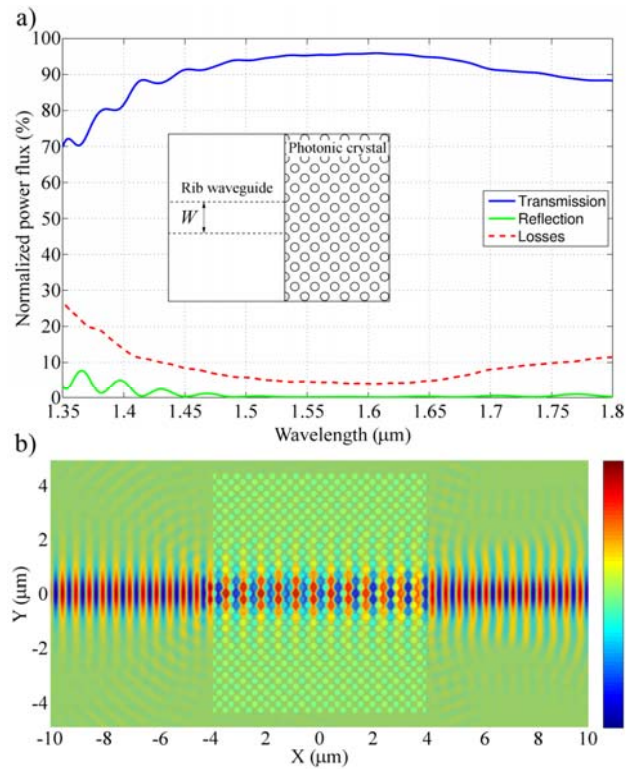

FIG. 4. (Color online) (a) Transmission (blue solid line), reflection (green solid line) and losses (red dashed line) spectra of the supercollimating structure, calculated with the FDTD method in 3D. The inset shows a top view of the device at the waveguide-PhC interface (b) Steady-state amplitude of the magnetic field $H_z$ at a wavelength of 1.55 μm in a top view of a cut in the middle of the device.